\documentclass[ ]{article}
\usepackage{cite}
\usepackage{amsmath,amssymb,amsfonts}
\usepackage{algorithmic}
\usepackage{graphicx}
\usepackage{textcomp}
\usepackage{xcolor}

\def\BibTeX{{\rm B\kern-.05em{\sc i\kern-.025em b}\kern-.08em
    T\kern-.1667em\lower.7ex\hbox{E}\kern-.125emX}}
    
\usepackage{multirow}
\usepackage{subfigure}
\usepackage{url}
\usepackage{enumitem}
\usepackage{cite} 
\usepackage{epstopdf}
\begin{document}

\title{Performance Analysis of Multipath BGP}

\author{
Jie Li\thanks{Jie Li is supported by China Scholarship Council (CSC) with grant no.\,201406060022.}\\
University College London\\
London, United Kingdom \\
\url{jie.li@cs.ucl.ac.uk}
\and
Shi Zhou\\
University College London\\
London, United Kingdom \\
\url{s.zhou@ucl.ac.uk}
\and
Vasileios Giotsas\\
Lancaster University\\
Lancaster, United Kingdom \\
\url{v.giotsas@lancaster.ac.uk}
}

\maketitle

\begin{abstract}
Multipath BGP (M-BGP) allows a BGP  router to install multiple `equally-good' paths, via  parallel {inter-domain} border links, to   a destination {prefix}.
M-BGP differs from the multipath routing techniques in many ways, e.g.\, M-BGP is only implemented at border routers of Autonomous Systems (ASes); and while it shares traffic to different IP addresses in a destination prefix via different border links, any traffic to a given destination IP always follows the same border link. 
Recently we studied Looking Glass data and reported the wide deployment of M-BGP in the Internet; in particular, Hurricane Electric (AS6939) has implemented over 1,000 cases of M-BGP to hundreds of its peering ASes.

In this paper, we analyzed the performance of M-BGP. We used  RIPE Atlas to send traceroute probes to a series of destination prefixes through Hurricane Electric's border routers implemented with M-BGP.   
We examined the distribution of Round Trip Time to 
each probed IP address in a destination prefix and their variation during the measurement. 
We observed that the deployment of M-BGP can guarantee stable routing between ASes and enhance a network's resilience to traffic changes.
Our work provides insights into the unique characteristics of M-BGP as an effective technique for load balancing. 

%Multipath BGP (M-BGP) allows a BGP router to install multiple 'equally-good' paths, via parallel inter-domain border links, to a destination prefix. M-BGP differs from the multipath routing techniques in many ways, e.g. M-BGP is only implemented at border routers of Autonomous Systems (ASes); and while it shares traffic to different IP addresses in a destination prefix via different border links, any traffic to a given destination IP always follows the same border link. Recently we studied Looking Glass data and reported the wide deployment of M-BGP in the Internet; in particular, Hurricane Electric (AS6939) has implemented over 1,000 cases of M-BGP to hundreds of its peering ASes. 

%In this paper, we analyzed the performance of M-BGP. We used RIPE Atlas to send traceroute probes to a series of destination prefixes through Hurricane Electric's border routers implemented with M-BGP. We examined the distribution of Round Trip Time to each probed IP address in a destination prefix and their variation during the measurement. We observed that the deployment of M-BGP can guarantee stable routing between ASes and enhance a network's resilience to traffic changes. Our work provides insights into the unique characteristics of M-BGP as an effective technique for load balancing.


\end{abstract}
\section{Introduction}
\label{Introduction}

The Border Gateway Protocol (BGP)~\cite{RFC4271} is the de-facto external gateway protocol for inter-domain routing. When a BGP router learns multiple paths to a destination IP prefix, it applies a ranking algorithm to select the best path~\cite{RFC4271}. BGP allows network operators to define their own policies on how to select the best paths, meaning that border routers can apply distinct and independent from each other routing policies. While BGP policy-based routing allows flexible route selection, it hinders the predictability of routing decisions -- especially without direct access to BGP configurations.  

By default, BGP selects a single best path to a destination. If two or more paths are equally good in terms of the configurable BGP attributes, BGP breaks ties using metrics such as the age of a path or the ID of the neighboring BGP router from which a path was received. Nonetheless, using multiple equivalent paths has the potential to improve both the performance and resilience of the routing system.  

{\em Multipath BGP} (M-BGP) has been introduced to enable load sharing between inter-domain paths of equal cost. 
Specifically, when multiple equally good eBGP (external BGP) paths are learned from the \textit{same} peering AS, and all the first six attributes of the BGP decision process (LocPref, AS path, Origin, MED, eBGP/iBGP, and IGP metric) have the same values, instead of applying last-resort tie-breaker, M-BGP installs all tied paths as active paths to the corresponding destination.
M-BGP is today supported by most major router vendors, including  Juniper~\cite{juniper-mbgp},  Cisco~\cite{CISCO},  and  Huawei~\cite{huawei-ecmp}.

Most load balancers are deployed in intra-domain routers,
since managing traffic within a single routing domain avoids the complexities introduced by the contractual relationships among ASes~\cite{Valera2011MBGP}.
Such load balancers are predominately per-flow or per-packet~\cite{Augustin2011TON, Vermeulen2018IMC}. 
%and those discussed in \cite{Singh2015IEEECST, Qadir2015IEEECST}) 
In contrast, M-BGP establishes multipath routing on border routers and while load sharing is typically applied on a per-flow basis, it only pertains to the subset of destination IP prefixes that can be reached by equally good paths received over different eBGP sessions. 
%Most M-BGP implementations allow by default load sharing over paths received from the same peering AS. However, many vendors allow this requirement to be relaxed and establish M-BGP over paths with different AS sequence, provided that they have equal AS path length~\cite{arista-ecmp, citrix-ecmp}.  

The increasing popularity of direct peering over IXPs to bypass transit providers and reduce path lengths has led to denser inter-domain connectivity at the edge of the network~\cite{gill-flattening-2008}, and therefore increases the potential benefits of M-BGP.
However, the extent of M-BGP deployment and its actual impact on AS paths is largely unexplored. M-BGP is still an optional function for inter-domain load sharing, and since it does not alter BGP updates, detecting its use needs to rely on data-plane measurements unless we have direct access to the configuration of border routers. 
Additionally, using traceroute data to determine load-balanced inter-domain links is non-trivial due to the challenges in accurately mapping inter-domain borders~\cite{yeganeh2019} and the number of measurements that need to be issued~\cite{Vermeulen2020NSDI}.
To provide a first analysis of M-BGP, we recently presented a methodology in~\cite{Li2020TMA} to measure the deployment of M-BGP in Hurricane Electric (HE, AS6939) by utilizing data from a set of BGP Looking Glass servers, and demonstrated some basic types of M-BGP deployment with traceroute data over the RIPE Atlas platform~\cite{RIPE2015IPJ}.

% has been a lack of researches on M-BGP for several reasons. Firstly, M-BGP is still an optional function for inter-domain load sharing. Secondly, the algorithm for M-BGP implementation should be compatible with existing BGP semantics and BGP routers within a network~\cite{Valera2011MBGP}. Thirdly, it is difficult to measure M-BGP from massive traceroute data, due to the challenge to accurately mapping inter-domain borders \cite{yeganeh2019}. Despite these challenges, we presented a methodology in a recent work of ours \cite{Li2020TMA} to measure the deployment of M-BGP in  Hurricane Electric (HE, AS6939) with data from Looking Glass, and demonstrated some basic types of M-BGP deployment with traceroute data. 

This paper extends our work in~\cite{Li2020TMA} by conducting performance analysis of M-BGP. 
Our results indicate that the deployment of M-BGP indeed guarantees stable routing performance between ASes and enhances a network's resilience to traffic changes. To be specific, when facing with traffic changes, either the routing between ASes remains stable or only one border link experiences increase of delay instead of all the border links, no matter the border links have the same bandwidth or not.  
Our results also suggest that the deployment of M-BGP can help networks deliver different types of traffic via different border links with rather stable performance. 
Our work contributes as the first attempt on studying the performance of M-BGP. Our study provides insights into the routing dynamics, the performance and the unique characteristics of M-BGP as an effective technique for load balancing.

The rest of the paper is organized as follows: In Section \ref{M-BGP_Definition} we provide our definition on M-BGP deployment, we then refine the method in~\cite{Li2020TMA} to a two-phase methodology (Section \ref{M-BGP_Inference}), and in Section \ref{M-BGP_Deployment} we apply our methodology to a wide range of ASes and provide evidence on the wide deployment of M-BGP. 
Then we analyze the theoretical benefits of M-BGP for Internet routing in Section \ref{Methodology_Theory}  and present empirical analysis based on traceroute data and Round Trip Time (RTT), with focus on Hurricane Electric in Section \ref{Methodology_Empirical}. We show three typical cases as case studies and examine the performance of M-BGP according to the distribution and variation of link delays for each destination IP during the measurement in Section \ref{Cases}. We discuss some related works in Section \ref{RelatedWorks} and conclude the paper in Section \ref{Conclusion}.  

\section{M-BGP Deployment in the Internet}
\label{M-BGP}

\subsection{Definition of M-BGP Deployment}
\label{M-BGP_Definition}

\begin{figure}
    \centering
    \subfigure[Normal routing]{
    \includegraphics[width = 0.65\textwidth]{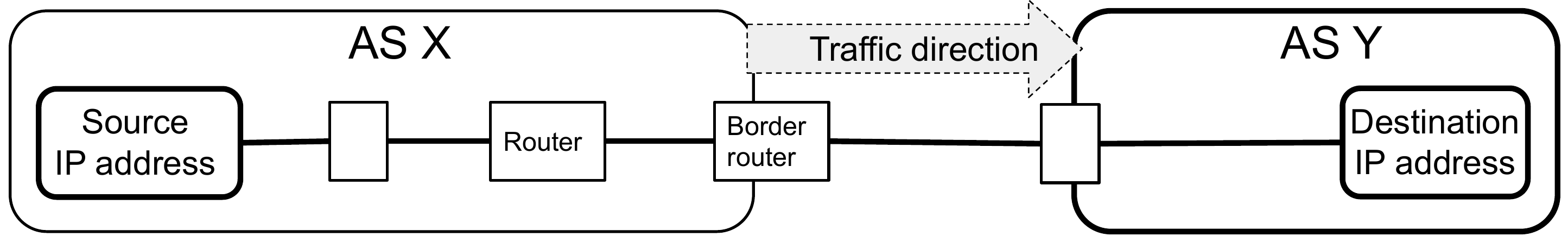}
    \label{fig:Examples-a}}
    \subfigure[Multipath routing]{
    \includegraphics[width = 0.65\textwidth]{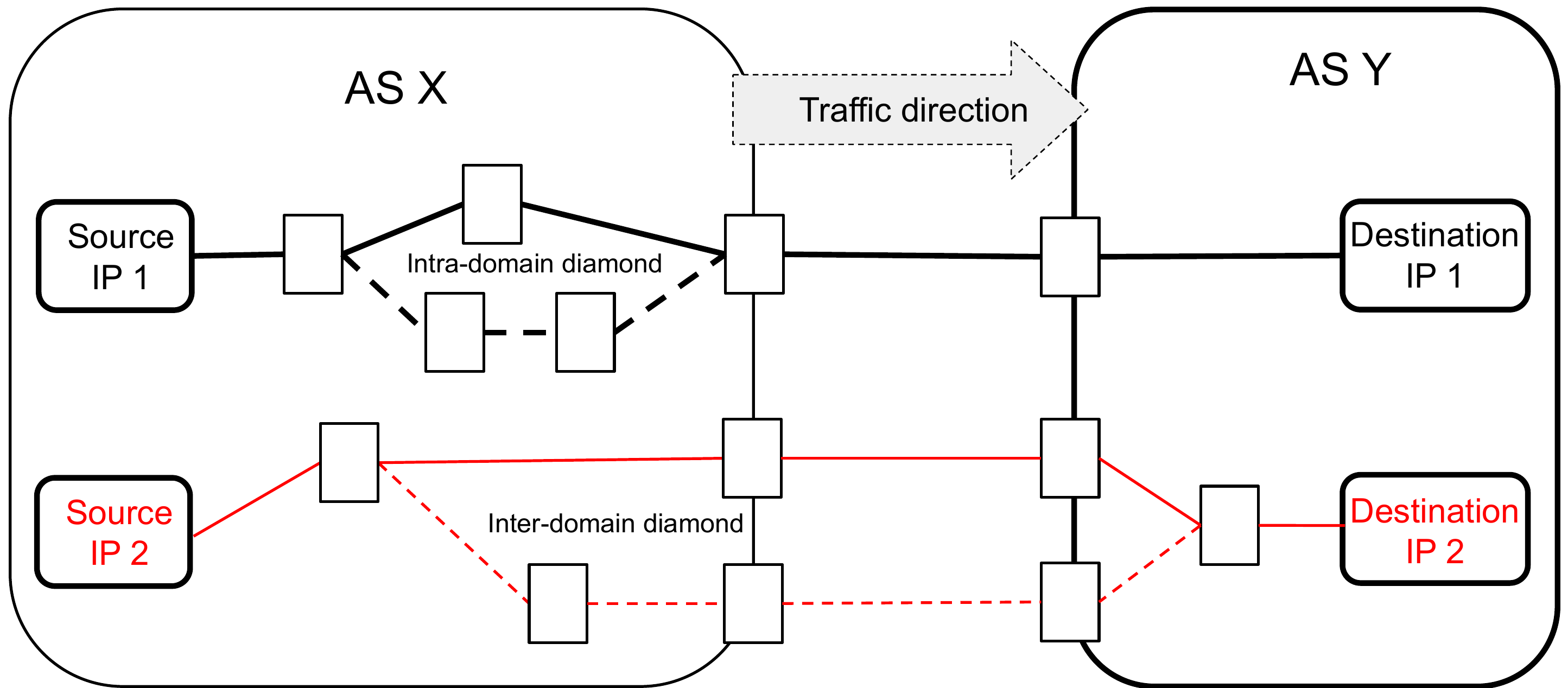}
    \label{fig:Examples-b}}
    \subfigure[Multipath BGP: topology map ]{
    \includegraphics[width = 0.65\textwidth]{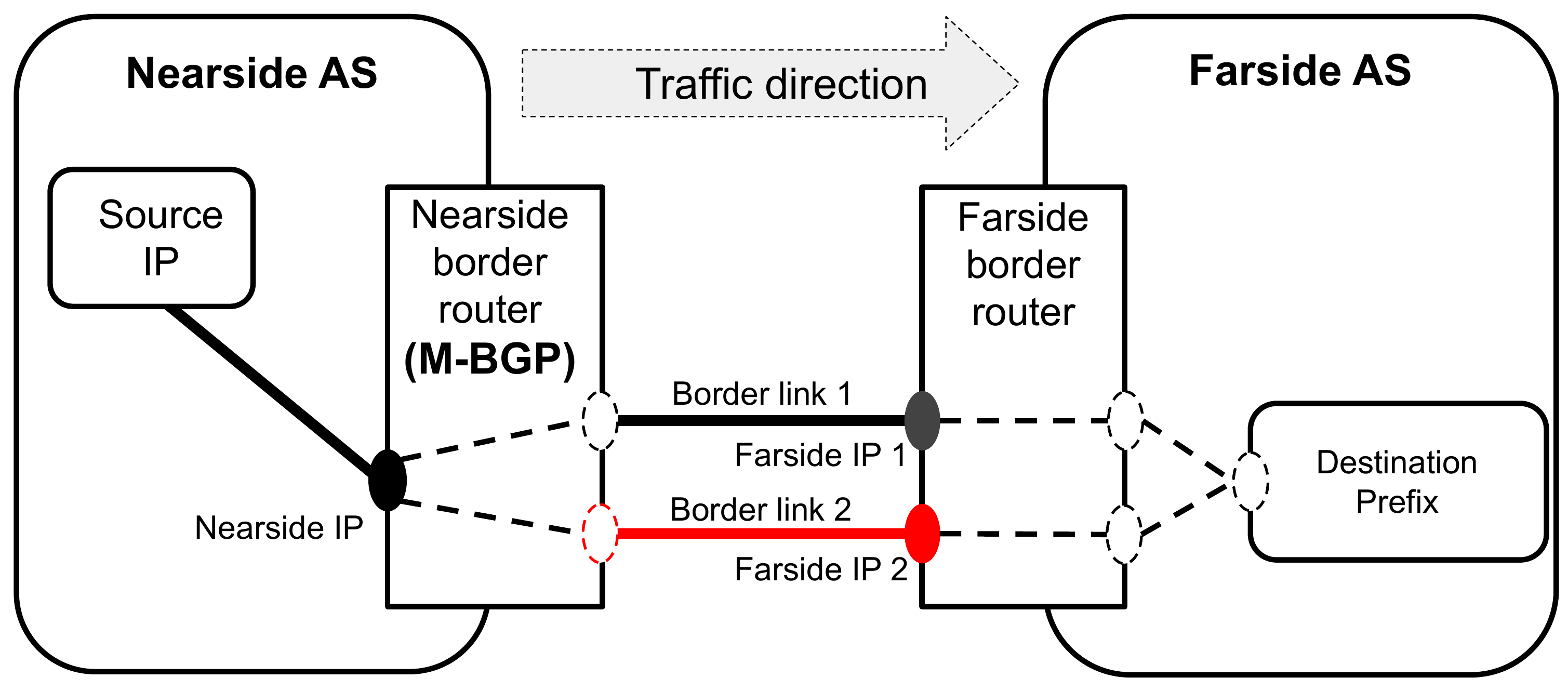}
    \label{fig:Examples-c}}
     \subfigure[Multipath BGP: traffic map ]{
    \includegraphics[width = 0.6\textwidth]{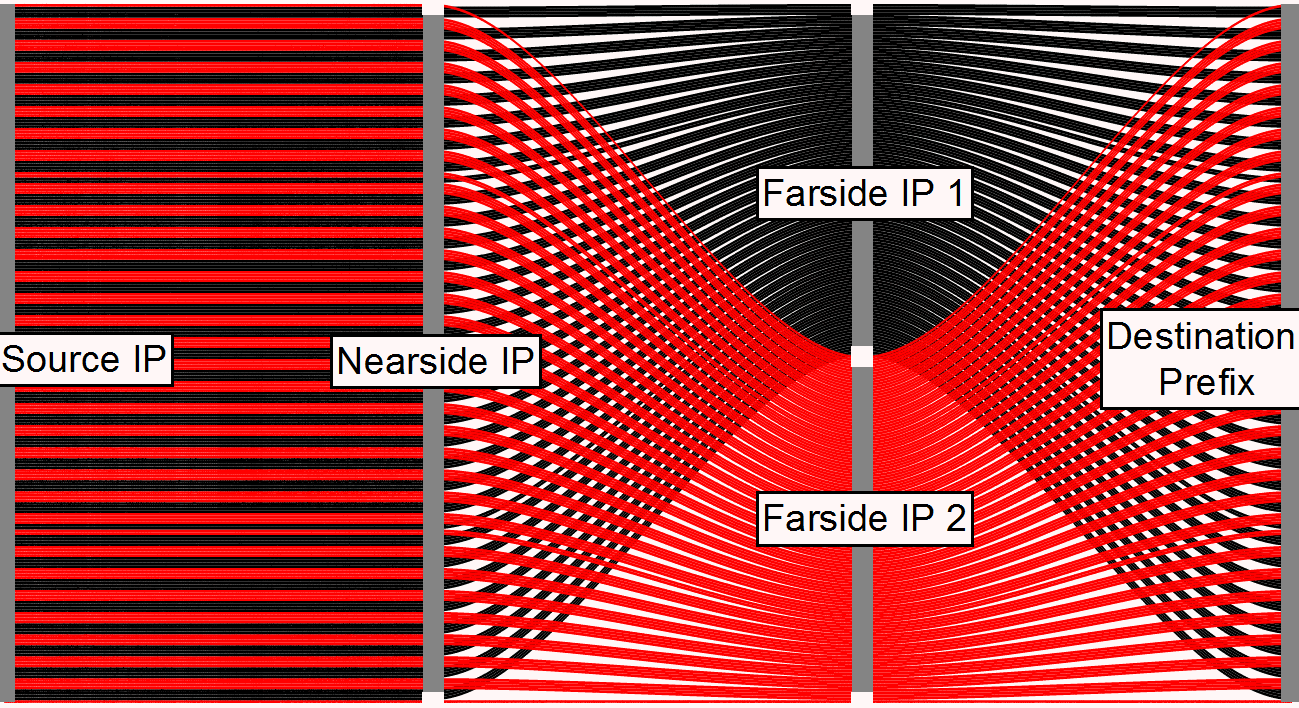}
    \label{fig:Examples-d}}
    
   \caption{Illustrative examples. (a) Normal routing, where a single path is used for routing between a source IP address and a destination IP address. (b)~Multipath routing, where multiple routing paths are used between a source IP address and a destination IP address -- the paths may diverge and merge within the same AS forming an intra-domain `diamond' \cite{Augustin2011TON, Vermeulen2020NSDI}, or cross AS borders forming an inter-domain diamond. (c) and (d)~Multipath BGP (M-BGP), where the Nearside Border Router uses multiple Border Links to share traffic flows to different IP addresses in the Destination Prefix while using a single, fixed path for each destination IP.}
   \label{fig:Examples}
\end{figure}

\begin{figure*}[!t]
    \centering
     \includegraphics[width= 1 \textwidth]{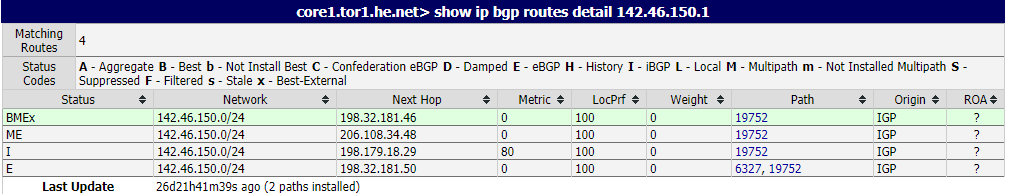}
    \caption{Example of LG response to the command of {\tt show ip bgp routes detail}}
    \label{fig:Responses}
\end{figure*}

Consider a traffic flow transiting from a Source IP address in a \textit{Nearside AS} ($AS_{near}$) to a Destination IP address in a \textit{Farside AS} ($AS_{far}$), as shown in Figure~\ref{fig:Examples}.
The two ASes can be connected by one or more \textit{Border Links} ($\mathcal{L}$). A \textit{Border Link} is a layer-3 interconnection between a \textit{Nearside Border Router} ($R_{near}$) and a \textit{Farside Border Router} ($R_{far}$). In a traceroute path, a \textit{Border Link} can be identified as two consecutive IP addresses that are mapped to different ASes, where the \textit{Nearside IP} and the \textit{Farside IP} are ingress interfaces of the two border routers. 

In the example of Figure \ref{fig:Examples-a}, there is only one \textit{Border Link} connecting the two peering ASes. $R_{near}$ installs only a single best route to the Destination IP, such that all traffic to the IP address follows the same border link.

When there are multiple Border Links connecting the two peering ASes, some of these links can be utilized for multipath routing as shown in Figure \ref{fig:Examples-b}, to split the traffic between the same source and destination IPs over the two alternative links. Such type of load sharing leads to paths that contain inter-domain `diamonds', namely path segments that have the same start and end IP hops, but different IPs in-between, and these path segments cross inter-AS boundaries~\cite{Augustin2011TON, Vermeulen2020NSDI}.

If there are more than one Border Links  between the same $R_{near}$ and $AS_{far}$, $AS_{near}$ can implement M-BGP at $R_{near}$ for a given Destination Prefix ($d$) (see Figure \ref{fig:Examples-c}), such that traffic flows to different IP addresses in $d$ are shared between the Border Links. 
 
%
 
%  {\color{red}
We use $<AS_{near},\, R_{near},\, AS_{far},\, d>$, a 4-parameter tuple, to denote a unique  case of M-BGP deployment. The tuple does not include $\mathcal{L}$, $R_{far}$, or the source of traffic because: $\mathcal{L}$ and $R_{far}$ can be determined by the four parameters;   and $R_{near}$ applies the same M-BGP settings to all traffic to (all IP addresses in)   $d$ regardless of the source. 
%  }
%$R_{far}$ is not included in the tuple because $\mathcal{L}$ can connect to either one $R_{far}$ or multiple $R_{far}$ in $AS_{far}$.}

%The source of traffic is not included in the tuple because $R_{near}$ applies the same M-BGP settings to all traffic to (any IP address in)   $d$ regardless of the source. 
%
For convenience, in this study we consider traffic flows starting in $AS_{near}$ and ending in $AS_{far}$, but the source of traffic can be outside of $AS_{near}$  and $d$ can be outside of $AS_{far}$ -- indeed they can be anywhere on the Internet as long as the traffic arrives at $R_{near}$ and traverses into $AS_{far}$.

If $AS_{near}$ and $AS_{far}$ are peering at an IXP, the M-BGP tuple does not need to include the IXP because IXP is `transparent' in BGP routing, i.e. the existence of IXP does not affect the function and deployment of M-BGP~\cite{RFC7947}. 

There are flexible ways to deploy M-BGP. For example, $AS_{near}$ can deploy M-BGP at different $R_{near}$ for the same $d$; or it can deploy M-BGP at the same $R_{near}$ for different $d$. All of these are considered as different cases of M-BGP deployment as they have different tuples. 
%Please note that no cases discussed in this paper fall into the second category.
%\textbf{Note that the cases studied in this paper don't include cases with those deployed at the same Nearside Border Router for different destination prefixes within the Farside AS.}

\subsection{Inferring M-BGP Deployment}
\label{M-BGP_Inference}

\subsubsection{Looking Glass (LG) Server Data}
\label{M-BGP_Inference_LG}

The definition of M-BGP given in Section \ref{M-BGP_Definition} indicates that the key to inferring M-BGP deployment is to locate the border routers of ASes. So far, a number of methods (e.g.~\cite{Luckie2016IMC, Marder2016IMC, Marder2018IMC}) have been proposed to map AS borders from traceroute data. However, even the state-of-the-art method, bdrmapIT \cite{Marder2018IMC},  can lead to erroneous border identification \cite{yeganeh2019}.

To alleviate this issue, we utilize Looking Glasses (LG) as a direct and reliable source of information on M-BGP deployment. They allow to query directly the BGP configuration and routing table of border routers. We have complied a list of 1,848 ASes with LG servers from data provided by  BGP Looking Glass Database \cite{BGPLGDatabase} and PeeringDB API \cite{PeeringDBAPI}. The next sections introduce a two-phase method to identify M-BGP deployment with LG data.

\subsubsection{Obtaining List of Peering ASes}
\label{M-BGP_Inference_PeeringAS}

As the first phase to identify M-BGP deployment, we  query each AS' border routers with the command {\tt show ip bgp summary} to obtain the AS' peering ASes at each border router. 
The command returns a summary table with the AS numbers of the BGP neighbors and the addresses of the remote IP interfaces through which the BGP session is established. %The addresses of the remote IP interfaces represent \textit{Border Links} in our study. %
%Figure \ref{fig:AS6939BGPSummary} is an example table  from the border router {\tt core1.tor1.he.net (tor1)}, listing partial information about the peering ASes at this router. 
In the summary table, some peering ASes are connected via multiple neighbor addresses, and these peering ASes are very likely to be deployed with M-BGP, because multiple next-hops is the condition for {\em tied} multipaths before M-BGP is activated.

\subsubsection{Identifying  M-BGP Deployment}
\label{M-BGP_Inference_Identification}

The second phase is to query each border router using command {\tt show ip bgp routes detail <IP address>} and identify the deployment of M-BGP. For each peering AS connected to a border router, we obtain a list of announced prefixes with data provided by RouteViews \cite{RouteViews}.
%Due to space constraint, this paper only focuses on IPv4 prefixes of length /24. 
%Then the parameter {\tt IP address} is set as {\tt x.x.x.1} for each destination prefix because queries to all the addresses in the same prefix should return the same routing table. 
Then we use one IP address in each prefix as the parameter for the command because queries to all the addresses in the same prefix should return the same routing table.

\begin{table}[t]
    \centering
    \caption{M-BGP deployment in the Internet.}  
    \label{tab:AS-M-BGP}

    \begin{tabular}{r|l|r|r|r}
    \hline
    &        & \multirow{4}{0.4 in}{\begin{tabular}[r]{@{}r@{}} \# of \\ {M-BGP} \\ Cases \end{tabular}} &  \# of     & \# of\\  
    AS       &    AS  &   & Peering ASes    & {Border Routers}\\  
    Number  &   Name &      & ({with M-BGP}     & ({with M-BGP}  \\ 
              &        &            & /{total})     & /{total})  \\                  
    \hline
    \multicolumn{5}{c}{IPv4}  \\  \hline
    6939    & HE        & 1,088 & 611/5,868 & 69/112    \\
    9002    & RETN      & 155   & 108/1,547 & 51/130    \\
    20764   & RASCOM    & 27    & 23/858    & 6/27      \\
    196965  & TechCom   & 24    & 15/36     & 2/2       \\
    22691   & ISPnet    & 3     & 3/24      & 1/7       \\
    3216    & VimpelCom & 2     & 2/770     & 2/16      \\
    12303   & ISZT      & 2     & 2/59      & 1/2       \\ 
    48972   & BetterBe  & 2     & 1/9       & 2/4       \\

    \hline
    
    \multicolumn{5}{c}{IPv6}  \\  \hline
    6939    & HE        & 300   & 146/3,880 & 35/112    \\
    9002    & RETN      & 45    & 23/926    & 24/130    \\
    48972   & BetterBe  & 2     & 1/6       & 2/4       \\
    \hline
    \multicolumn{5}{l}{HE: Hurricane Electric; RASCOM: CJSC RASCOM; }  \\
    \multicolumn{5}{l}{VimpelCom: PJSC VimpelCom}  \\
    \end{tabular}
\end{table}

Figure~\ref{fig:Responses} shows an example response to the command from {\tt core1.tor1.he.net}, a border router of HE.
The figure shows that two paths are  installed towards the destination prefix. They are labelled with status codes of ``M'' and ``E'', meaning they are multipath learned via external BGP. They also have same values for metrics including LocPref, Weight, Path, Origin, and Metric. This indicates that HE has deployed M-BGP to AS19752 at this border router. %The IP addresses 108.32.181.46 and 206.108.34.48

If a prefix in a peering AS is identified as having M-BGP deployment at a border router, we record this as an M-BGP case and the query  goes to the next peering AS. As a proof of concept, we do not aim to identify all the prefixes with M-BGP deployment within a peering AS. If all the prefixes in the peering AS are queried and no M-BGP deployment is identified, the query also goes to the next peering AS. When all the peering ASes connected to a border router are queried, the query goes to the next border router. 

\subsection{M-BGP Deployment in the Internet}
\label{M-BGP_Deployment}

% {\color{red}
We have applied the method to 2,709 ASes, and identified M-BGP cases deployed by 8 ASes on IPv4 and by 3 ASes on IPv6. Table \ref{tab:AS-M-BGP} lists the information about these ASes, ranked according to their numbers of identified M-BGP cases and AS number as tie-breaker.  The table shows that HE has deployed much more cases than the other ASes. Because HE is also a top rank ISP network, we focus on HE  to analyze how M-BGP performs as a load sharing technique.
% }

\section{Performance Analysis of M-BGP Deployment}
\label{Methodology}

Although M-BGP has been widely deployed in the Internet,   there is no study in literature on the performance of M-BGP. 
Here we present an empirical study on M-BGP performance based on  traceroute measurements. 

\subsection{Expected Benefits of M-BGP for Internet Routing}
\label{Methodology_Theory}

When M-BGP is deployed, multiple paths are learned, installed and shared for traffic load to a destination prefix, which should bring benefits to routing performance.
For example, M-BGP shares traffic load over multiple border links, which should reduce congestion and improve network resilience against link failure and sudden traffic surge.  

Comparing to multipath routing, which is another load sharing technique where traffic  to a same IP address follows different paths, M-BGP has a distinct advantage. That is, although M-BGP uses different border links for traffic to a destination prefix, it ensures that all traffic to any IP address in the prefix always follows the same border link. 
%This reflects that BGP is designed to choose paths for prefixes instead of IPs, and ensures a fixed routing path for a specific pair of source and destination IPs. 
%
This is significant. While multipath routing may disrupt the sequential transmission of data packets from  source to  destination, M-BGP can guarantee the sequential transmission and therefore lead to stable performance at the TCP level.

\subsection{Empirical Analysis on M-BGP Performance}
\label{Methodology_Empirical}

Ideally, we should  obtain traffic data on border links before and after a deployment of M-BGP for performance analysis. However, it is unpractical to known or predict the timing of M-BGP deployment. %Moreover, continuous monitoring of a network's traffic data may be problematic to the target network. 

%Ideally, we can  analyze the difference in links' performance (e.g. traffic volume) before and after M-BGP is deployed. However, it is difficult to know (or even predict) when and where M-BGP is (or will be) deployed, hence unable to obtain the performance data. Another option is to monitor links already deployed with M-BGP and capture data when M-BGP is no longer deployed. However, continuous monitoring is very likely forbidden because it may cause unnecessary burden or even damage to the target networks. 
% {\color{red}
Here we propose a solution based on active traceroute probing using RIPE Atlas \cite{RIPE2015IPJ}. 
We used default settings of RIPE Atlas , e.g.\,ICMP messages and Paris traceroute variation 16.
We selected 15 M-BGP cases in HE for performance analysis, because  border links of these cases can be observed in traceroute probes sent from  RIPE Atlas probes located in HE to IP addresses within their Destination Prefixes. 
%
%If a case has a Sole-Destination Prefix, it only needs to satisfy rule (1); otherwise it should satisfy both rules. 
%
%We selected 15 M-BGP cases for analysis, including 11 cases where with only Destination Prefix and 4 cases with both Destination Prefix and Non-Destination Prefixes. 
%

For each M-BGP case, firstly we   sent traceroute probes to the first 100  IP addresses in the Destination Prefix every 15 minutes for 24 hours, i.e. each IP was probed $4 \times 24 = 96$  times. %
%we run traceroute measurements to Destination Prefix and then compare with those to the Non-Destination Prefixes. 
%
%
%We analyzed the performance of M-BGP by studying the Round Trip Time (RTT) values.
%
Secondly, we calculated the Round Trip Time (RTT) value at each IP hop.
%because traceroute returns at most three RTT values for each responsive interface. %
Then, we calculated the \emph{delay on a border link}, which is the difference between the RTT values of the Nearside IP and the Farside IP of the border link.
The delay consists of the (round trip) transmission time on a border link and the message processing time at $R_{far}$. %
We probed the first one hundred IP addresses in each prefix due to the limit set by RIPE Atlas on simultaneous measurements for each account. The 15-minute interval is to ensure no interference between two consecutive probes to the same destination IP.
For comparison purpose, for each M-BGP case, we also sent traceroute probes to a \textit{Non-Destination Prefix}, where only one of the border links is traversed. 
%
%*** For a given M-BGP case $<AS_{near},\, R_{near},\, AS_{far},\, d>$,  other prefixes in the same $AS_{far}$ are called \textit{Non-Destination Prefixes}.
%
%***and (2) traceroute probes to any IP address in a Non-Destination Prefix should traverse only one of the border links to the Destination Prefix.}
%

%One may argue that the border link delay is  for the round trip instead of one way delay, and the forward path may be different from the reverse path. We believe the effect of round trip delay and forward-reverse path difference has been mitigated in our method from two aspects. 

%The transmission time is round trip time but still on the border link, and the processing time is closely related to the traffic on the link. Second, our 
%study is based on statistical analysis with plots of 25\%, 50\% (median) and 75\% delay values.
% }

\begin{figure}[!t]
    \centering
    \subfigure[Delays on Border Link~1]{
    \includegraphics[width = 0.8\textwidth] {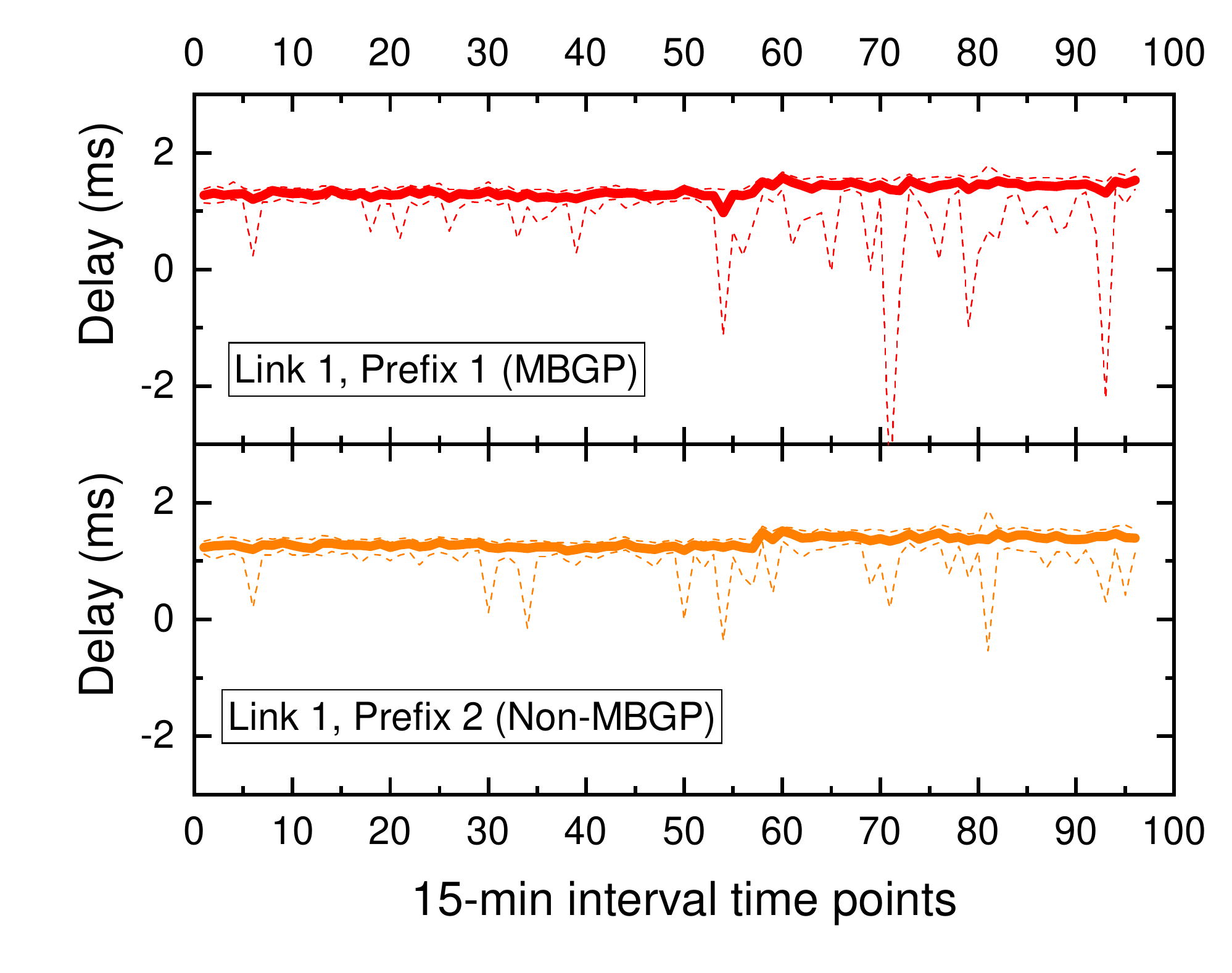}}
   \subfigure[Delays on Border Link~2]{
    \includegraphics[width = 0.8\textwidth] {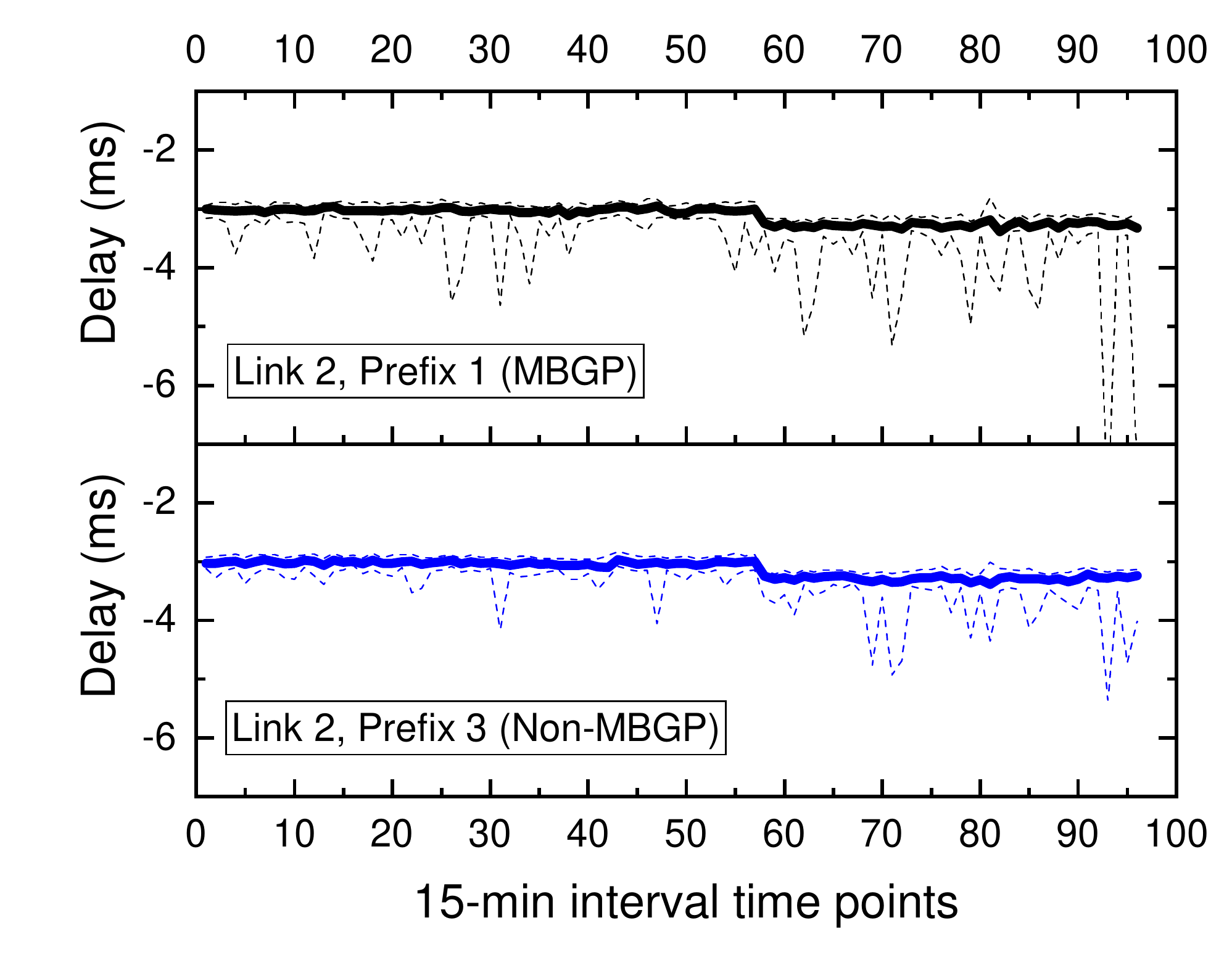}}
    \caption{Case 1. Delays on the two border links. Both links are used for traffic to the Destination Prefix (Prefix 1). Link~1 and Link~2 are used for traffic to two Non-Destination Prefixes (Prefix 2 and Prefix 3), separately and respectively.
    }   
    \label{fig:case1}
\end{figure}

\begin{figure}[!t]
    \centering
    \subfigure[Delays on the two border links, where the inset shows   delays on Link~1 for Time Points 1--25.]{
    \includegraphics[width = 0.85\textwidth] {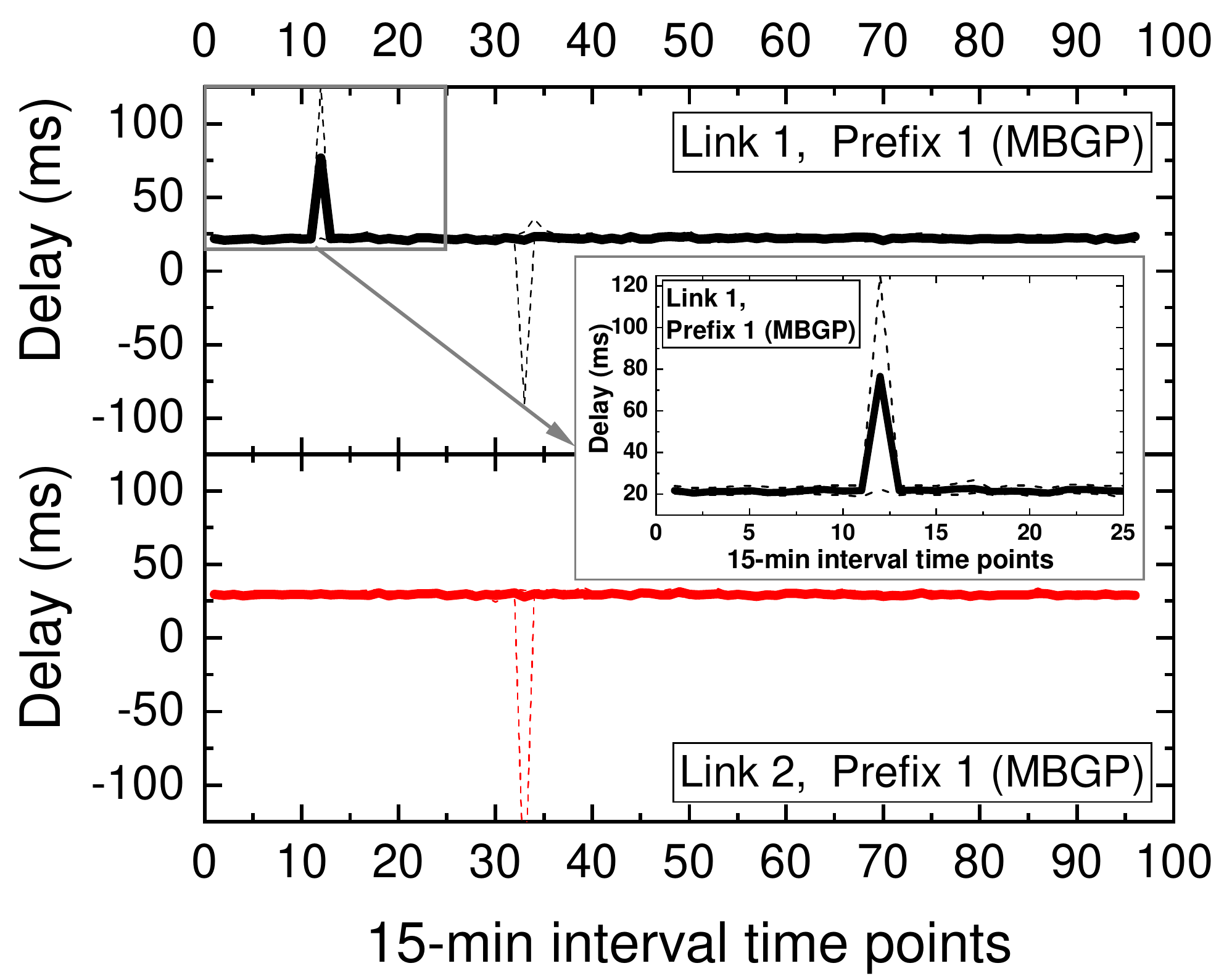}
    \label{fig:case2-a}}
    \subfigure[Distributions of delays on Link 1 and Link 2 at Time Point 12 and Time Point 20.]{
    \includegraphics[width = 0.8\textwidth] {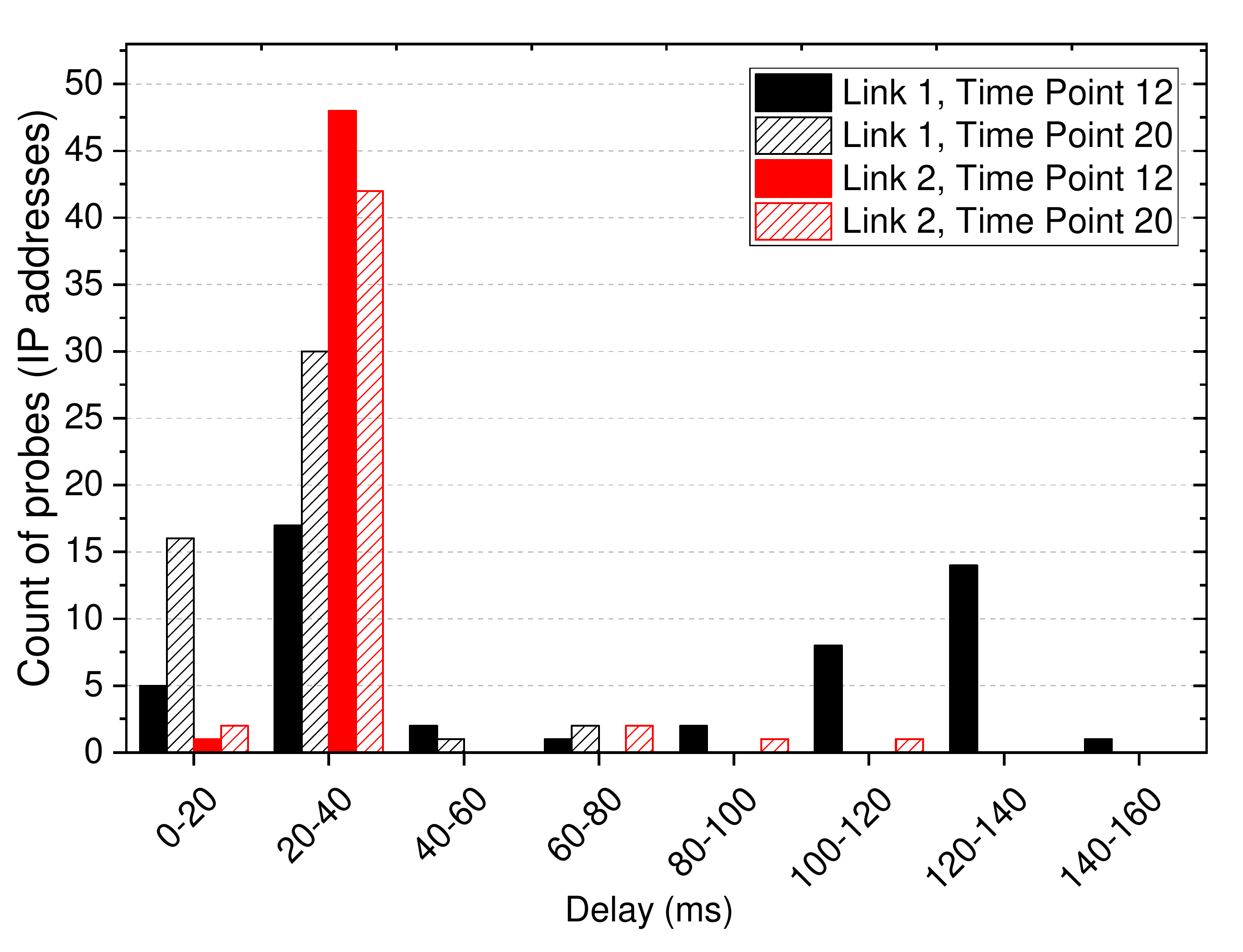}
    \label{fig:case2-b}}
    \caption{Case 2.}   
    \label{fig:case2}
\end{figure}

\section{Case Studies on M-BGP Performance}
\label{Cases}

%In Section \ref{Methodology_Empirical}, we selected 15 cases for performance analysis.  
Due to limit of space, this paper presents three case studies chosen from the 15 M-BGP cases that we measured above. %s among them as case studies, and omits the detailed IP information in each case. 
In Figs. \ref{fig:case1}-\ref{fig:case3},  border link delays are plotted at 25th (dashed line), 50th (i.e. the median, solid line), and 75th (dashed line) percentiles 
in increasing order 
%for each prefix on each link 
at each time point. The bandwidth of each border link is provided by PeeringDB \cite{PeeringDB}.

\subsection{Case 1}
Figure \ref{fig:case1} plots the result for Case 1, where M-BGP is deployed at HE's Border Router {\tt core1.hkg1.he.net}  ({\tt hkg1}) to AS10118 via two border links  with the same bandwidth.

%One Destination Prefix and two Non-Destination Prefixes are probed. Both links are used for the Destination Prefix, and each link is used for an Non-Destination Prefix.
We can observe that both border links experienced a change of delay 
%(i.e. the 15th hour) 
for traffic to both Destination and Non-Destination Prefixes at Time Points 57-60. 
%The delay on Link~1 increased, whereas the delay on Link~2 reduced by a similar amount. 
%
The change remained for the rest of the measurement, indicating a long-term  change happened to the networks
%to AS10118  
at that time. We also observe that after the  change, the delay on the links showed more fluctuation while their median (50th percentile) values still remained stable.
%, suggesting a rather stable routing performance for all the prefixes.  
% 
In this case, a long-term network change had a similar impact for M-BGP routing and none-M-BGP routing. 
%{\color{red}
%Although this case does not show significant better performance of M-BGP, it is still important to our study that the deployment of M-BGP guarantees the stable routing.
%}

\subsection{Case 2}
In Case 2, M-BGP is deployed at the same Border Router of HE as in Case 1 ({\tt hkg1}) but to a different Farside AS, AS20940,  via two border links (which of course are different from those in Case 1). 

% {\color{red}
Figure \ref{fig:case2-a} shows that although the two links had stable routing performance at most time points, 
%and a same fluctuation at Time Point 33 
Link~1 experienced a sharp increase of delay at Time Point 12, which did not occur on Link 2 at all. 
The inset in Figure \ref{fig:case2-a} shows the median of delay on Link~1 jumped, from the normal delay of 20ms, suddenly to 75ms at Time Point 12 and then immediately returned to normal at the next time point.

Figure \ref{fig:case2-b} plots the
frequency distributions of  delays on Link 1 and Link 2 (to different sets of IP addresses in the Destination Prefix)  at Time Point 12 (i.e. surge of Link 1 delay) and Time Point 20 (i.e. stable status), respectively. 
We can see that normally, as measured at Time Point 20,   delay on both links are mostly below 40ms. Whereas at Time Point 12, there was a surge of delay on Link 1, where traceroute probes to 23 IP addresses experienced more than 100ms delay on Link 1. 
Such a sharp increase of traffic delay on Link 1 was likely caused by a sudden rise of traffic volume to these IP addresses allocated to Link 1 by M-BGP. 

Notably, there is no such delay on Link 2 at all at that same Time Point.
%
%In this case, Link~2 has a higher bandwidth (100G) than Link~1 (10G). It is possible Link 2 has also experienced increase of traffic but not shown in the results because of the high bandwidth. However, Figure \ref{fig:case2-b} shows that the median delay on Link 2 at Time Point 12 is even better than that at Time Point 20, suggesting no increase of traffic on Link 2. Therefore, 
%
The reason that Link 2 completely avoided this sharp increase of delay
%(shown in both Figure \ref{fig:case2-a} and Figure \ref{fig:case2-b})  
is  due to the M-BGP deployment, which routed traffic to different sets of IP addresses in the Destination Prefix via different border links. Thus, a surge of traffic to IP addresses allocated to one border link would have little impact on the routing performance of another border link.
% }    

\begin{figure}[!t]
    \centering
    \includegraphics[width =0.85\textwidth] {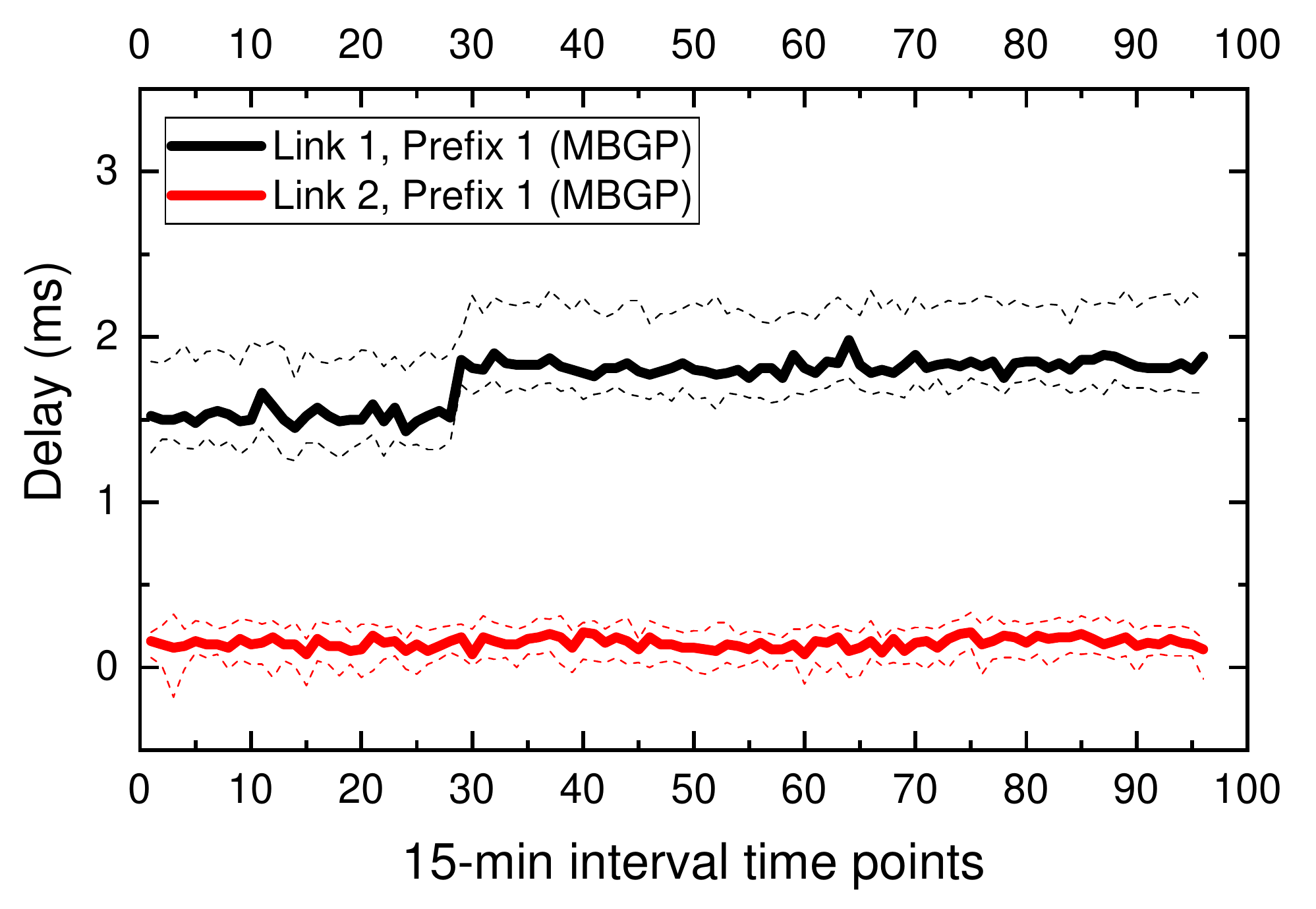}
    \caption{Case 3. Delays on two border links.}   
    \label{fig:case3}
\end{figure}

\subsection{Case 3}
In Case 3, M-BGP is deployed at HE's Border Router {\tt core1.sin1.he.net} ({\tt sin1}) to AS9930 via two border links. %Only the Destination Prefix is probed. 
Figure \ref{fig:case3} shows Link~1 consistently experienced higher delays and higher fluctuation than Link~2.  This is consistent with the fact  that Link 1 has a lower bandwidth (10G) than Link 2 (100G).  
%The reason for this can be either the links are different in physical characteristics, or the links are loaded with different volumes or even types of traffic. 

 The benefit of M-BGP deployment is shown at the time point 28 when there is a  significant and permanent increase of traffic delay on Link 1, possibly due to an increase of traffic to IP addresses that transit through Link 1; whereas such traffic increase has no effect on Link~2 whose link delay remained stable during the entire period of measurement.

Case 2 and Case 3 demonstrate that M-BGP allows a network operator to use different border links for  different  types of traffic to different IP addresses in the same destination prefix. If  destination IPs with more variable traffic loads are allocated to one link, then routing performance to other IPs transiting through other border links can be better protected and  guaranteed. Network operators may find this functionality useful, which can be conveniently implemented by M-BGP.

\section{Related Works}
\label{RelatedWorks}

\subsection{Multipath BGP}
\label{RelatedWorks_M-BGP}

To the best of our knowledge, the studies on M-BGP are limited in literature.
For example, Valera \textit{et al.}~\cite{Valera2011MBGP} explained the motivations to apply M-BGP and discussed some alternatives to M-BGP. A recent work of ours \cite{Li2020TMA} took Hurricane Electric as a case study, used Looking Glass data to infer the deployment of M-BGP, and analyzed some basic patterns of M-BGP deployment with traceroute measurement data. 
Therefore, while M-BGP has been supported by some major router vendors, we still need more knowledge about M-BGP. This paper contributes as the first attempt to understand the performance of M-BGP with analysis based on traceroute data. 

\subsection{Round Trip Time (RTT)}
\label{RelatedWorks_RTT}

Round Trip Time (RTT) has been widely studied in Internet research for different purposes. %For example, in %\cite{Ahmad2010ImpactIXP}, the authors analyzed the RTT values to help understand the impact of IXPs on the Internet routing performance. 
Some researches study the relation between RTT and routing patterns. For example, 
%Landa \textit{et al.} \cite{Landa2013IFIPNetworking} presented a model to analyze Internet RTT and its relationship to geolocation distance, and revealed large-scale routing information.
Javed \textit{et al.} \cite{Javed2013CCR} used the relative changes in RTT to study the root cause of path changes. 
Rimondini \textit{et al.}  \cite{Rimondini2013BGPRTT, Rimondini2014PAM} analyzed RTT measurement data, matched and correlated the BGP routing changes with RTT variations. Shao \textit{et al.} \cite{Shao2017ITC} presented an analysis framework to detect changes on RTT time series and to distinguish path changes due to routing protocols. 
%Castro \textit{et al.} \cite{Castro2014CoNEXT} relied on RTT values to detect remote peering networks. 
%Fanou \textit{et al.} \cite{Fanou2017Comput.Netw.} tracked the topological changes in Africa inter-domain routing with RTT analysis. 
%Formoso \textit{et al.} \cite{Formoso2018INFOCOM} explored the current state of the African Internet based on RTT obtained from a measurement campaign methodology. 
Mouchet \textit{et al.} \cite{Mouchet2020IEEEAccess} proposed to use infinite hidden Markov model for accurate representation of measured RTT time series from large scale traceroute data.

Some researches focused on the network delays with RTT data. Kotronis \textit{et al.} \cite{Kotronis2017IMC} conducted RTT measurements to study the selection of network relays.
Fontugne \textit{et al.} \cite{Fontugne2017IMC} deployed traceroute measurements, collected RTT data, and proposed several methods to detect and pinpoint delay anomalies in the Internet. 

Our work also uses RTT values but differs from the existing researches by providing preliminary analysis about the routing performance of M-BGP.

%Apart from the above topics, RTT has also been studied for prediction-related purposes. We refer readers to the survey paper by Mirkovic \textit{et al.} \cite{Mirkovic2018CST} for detailed introduction.

\section{Conclusion}
\label{Conclusion}

Following our recent work on inferring M-BGP deployment in the wild Internet, this paper reported our empirical measurement study on performance of M-BGP. 
%, including our methodology to infer M-BGP deployment cases, our traceroute measurement for the analysis, and three typical case studies as well. 
Our result supports the notion that the deployment of M-BGP can improve a network's resilience to changes and therefore enhance routing performance in general by sharing and separating traffic to IP addresses in a destination prefix.   This paper highlights the unique characteristics of M-BGP as an effective technique for load balancing.  

%We are currently working on identifying as many Destination Prefixes as possible within farside ASes. Such that we can conduct more traceroute measurements to obtain in-depth understanding about M-BGP. %Another line of future works is to expand the research to M-BGP cases deployed on IPv6 network.

\end{document}